\documentclass[aps,prd,
preprintnumbers,nofootinbib,superscriptaddress]{revtex4}
\usepackage[dvips]{graphicx}
\usepackage{caption}
\usepackage{subcaption}
\usepackage{bm,latexsym,amsmath,amssymb,amsfonts,color}
\usepackage{float}
\newcommand{\dn}{\operatorname{dn}}

\usepackage[normalem]{ulem}

\begin{document}

\title{On the robustness of the chiral soliton lattice and {the propagation of test quarks}}

\author{Fabrizio Canfora}
\email{fabrizio.canfora@uss.cl}
\affiliation{Centro de Estudios Cient\'{\i}ficos (CECS), Casilla 1469,
Valdivia, Chile}
\affiliation{Universidad San Sebasti\'{a}n, sede Valdivia, General Lagos 1163, Valdivia 5110693, Chile}
\author{Nicol\'as Grandi}
\email{grandi@fisica.unlp.edu.ar}
\affiliation{Departamento de F\'isica Dr. Emil Bose, Universidad Nacional de La Plata,
Casilla de Correos 67, La Plata, Argentina}
\affiliation{Instituto de F\'isica La Plata, Consejo Nacional de Investigaciones Cient\'ificas y T\'ecnicas,
Diagonal 113 esquina 63, La Plata, Argentina}
\author{Marcela Lagos}
\email{marcela.lagos@uss.cl}
\affiliation{Universidad San Sebastián, Avenida del Cóndor 720, Santiago, Chile}
\author{Luis Urrutia-Reyes}
\email{lurrutia2018@udec.cl}
\affiliation{Centro de Estudios Cient\'{\i}ficos (CECS), Casilla 1469,
Valdivia, Chile}
\affiliation{Universidad San Sebasti\'{a}n, sede Valdivia, General Lagos 1163, Valdivia 5110693, Chile}
\affiliation{Departamento de F\'{\i}sica, Universidad de Concepci\'{o}n,
Casilla 160-C, Concepci\'{o}n, Chile}
\author{Aldo Vera}
\email{aldo.vera@umayor.cl}
\affiliation{N\'ucleo de Matem\'atica, F\'isica y Estad\'istica, Universidad Mayor, Avenida Manuel Montt 367, Santiago, Chile}
\affiliation{Centro Multidisciplinario de F\'isica, Vicerrector\'ia de Investigaci\'on, Universidad Mayor, Camino La Pir\'amide 5750, Santiago, Chile}

\begin{abstract}
In this paper, we demonstrate the robustness of the chiral soliton lattice (ChSL) against a specific class of operators: the higher-order corrections in the 't Hooft large-$N_c$ expansion within the generalized Skyrme model coupled to Maxwell theory. By considering a suitable Ansatz adapted to describe topological solitons at finite baryon density in a constant magnetic field, the generalized Skyrme model coupled to Maxwell theory is reduced to the effective Lagrangian of the ChSL phase, which describes a lattice of domain-walls made of hadrons. One of the key points in this construction is the fact that even when the usual topological charge density vanishes, the presence of the Callan-Witten term in the topological charge density allows for a non-vanishing baryon number. In the present approach, rather than assuming a prescribed external magnetic field (as is usually assumed for the ChSL), we solve the fully coupled Skyrme-Maxwell field equations. Finally, we show how our formulation allows us to study the interaction of the ChSL with test quarks. In particular, we derive the exact analytical spectrum of the Dirac equation in the high-density limit, providing a microscopic characterization of the fermionic excitations within the inhomogeneous hadronic background provided by the ChSL. The comparison of the spectrum of the Dirac operator in the ChSL with that of the standard Dirac operator in a constant magnetic field reveals the fundamental role of both the quark-Skyrmion coupling and the hadronic profile in opening a gap and shifting the spectrum.
\end{abstract}

\maketitle

\newpage 

\section{Introduction} \label{sec-1}

The ChSL is a hadronic phase consisting of a lattice of aligned parallel domain-walls (two-dimensional topological solitons) separated at equal spatial intervals. This phase appears in many contexts, from condensed matter physics to high-energy physics \cite{Dzyaloshinskii}-\cite{Son2}. In recent years, the ChSL has attracted a lot of attention due to its role in quantum chromodynamics (QCD) under extreme conditions. In fact, it turns out that the ChSL is the ground state of QCD at finite baryon chemical potential in the presence of a critical homogeneous magnetic field \cite{Andersen:2014xxa}-\cite{Brauner:2019aid}. Beyond a sufficiently strong magnetic field, the ChSL becomes the energetically preferred hadronic phase, stabilized by the presence of a topological charge that emerges from the chiral anomaly. Relevant properties of the ChSL have been studied, such as the identification of the sector of the phase diagram where this phase can exist \cite{Brauner:2021sci}-\cite{Eto:2023wul}, its extension to spinning baryonic matter phases \cite{Nishimura:2020odq}-\cite{Chen:2022mhf}, the formation of topological solitons \cite{Schmitt1}-\cite{Vera:2024edv} and its relation to holography \cite{Sakai:2004cn}-\cite{Amano:2025iwi} (see also Refs. \cite{Yamada:2021jhy}-\cite{Harada:2006di}).

The ChSL can be described by the effective sine-Gordon theory (which admits topological multi-soliton solutions) in an external magnetic field and is supplemented by the Wess-Zumino-Witten term. Although this effective action describes the ChSL phase well, its derivation from quantum field theories is not yet completely clear. Obtaining the phase from first principles is the novelty of this article. In particular, the first question we will answer (affirmatively) is the following: Is it possible to construct the chiral soliton lattice even when the magnetic field is a dynamical field minimally coupled to the hadronic field? Indeed, a very important issue is to clarify what happens when the magnetic field is not an external field.

In the context of effective field theories, the Skyrme model describes the low-energy limit of QCD at leading order in the large-$N_c$ expansion \cite{Skyrme1}, \cite{Skyrme2}. Its topological soliton solutions are fermionic degrees of freedom recognized as baryons \cite{Witten}, \cite{ANW}, which arise due to the non-linear interaction between mesons (see Refs. \cite{WeinbergBook} and \cite{MantonBook}). This theory can be minimally coupled to Maxwell's electrodynamics to describe low-energy charged hadrons. In Ref. \cite{CW} (see also \cite{Piette:1997ny}), it was shown that when this coupling is performed, the usual expression for the topological charge density must be modified to ensure its conservation and maintain invariance under gauge transformations.

In this manuscript, we will derive the ChSL from the gauged Skyrme model. Conceptually, this construction complements previous works such as \cite{Brauner:2016pko}-\cite{Higaki}, which employ Chiral Perturbation Theory ($\chi$PT). While $\chi$PT organizes the most general low-energy mesonic Lagrangian as an expansion in derivatives and quark masses, the Skyrme framework provides a specific non-linear completion (incorporating the stabilizing four-derivative term) that admits a semiclassical interpretation motivated by the large-$N_c$ limit of QCD. In practice, both approaches offer a natural setting for incorporating higher-derivative subleading corrections, as well as for coupling the chiral background to spin-isospin Dirac fermions.

In order to perform such a derivation, here we will use a modification of the Ansatz presented in Refs. \cite{Alvarez:2017cjm}-\cite{Canfora:2022jmh}, which has allowed the construction of exact solutions that describe baryonic crystalline structures at finite volume in the gauged Skyrme and Yang-Mills theories. In particular, using the exponential representation of $SU(2)$ (being in this context the isospin global symmetry group), we will show the explicit form of the Skyrme field $U(x)$ in terms of a single one-dimensional soliton profile. Furthermore, by knowing the exact form of such a profile, it becomes possible to study analytically the propagation of test quarks in the ChSL background. The Dirac equation with a Skyrmion background has been addressed several times in the literature, motivated in part by the nucleon-meson interaction (in the limit where the nucleon is much heavier than the meson) in the context of nuclear matter \cite{Krusch1}-\cite{zhao}, but more broadly as a way to study the general interplay between fermions and topological solitons \cite{balachandran1}-\cite{Shnir}. However, due to the difficulty of this problem, analytical solutions could only rarely be obtained.

In this framework, we perform an explicit calculation of the fermionic excitations by deriving the exact analytical spectrum of the Dirac equation in the high-density limit, providing a clear picture of how the inhomogeneous hadronic background affects the underlying quark degrees of freedom. {Physically, the setup of this computation can be trusted until the hybrid-phase, where the temperature is near the confinement phase and a few quarks can propagate in the configuration while, at the same time, baryons still exist (at higher temperatures, the model is no longer valid). In this regime, we treat quarks as probe fermions in the two-flavor case, thus extending prior works cited above to an inhomogeneous and periodic Skyrmionic background.\footnote{While the fermionic spectrum may not be directly observable, we conjecture that such an analysis may provide microscopic input related to the electromagnetic, thermodynamic, and transport response of the ChSL. Furthermore, possible phenomenological applications include charge-dependent observables in heavy-ion collisions and cooling or transport processes in strongly magnetized compact stars.}

On the other hand, a natural question arises. It is well known that, in the ’t Hooft expansion (see \cite{Marleau:1989fh}-\cite{Gudnason:2017opo} and references therein), sub-leading corrections to the Skyrme model appear. Such corrections are extremely complicated and (at first glance) could destroy or, at the very least, substantially modify the neat analytic form of the chiral soliton lattice. Quite surprisingly, we will show that regardless of how many sub-leading terms are included in the action, the ChSL remains unchanged. In order to improve the clarity of the presentation, in the main text we will only quote the relevant results about the robustness of the ChSL, while the technical discussion will be included in the Appendix. 

This paper is organized as follows. In Sec. \ref{sec-2} we give a brief review of the gauged $SU(2)$-Skyrme model. In Sec. \ref{sec-3} we show how to construct exact topological soliton solutions at finite volume. Then, in Sec. \ref{sec-4} we derive the effective Lagrangian of the ChSL from the gauged Skyrme model using the Ansatz that allows the exact domain-wall configurations previously constructed. We explore the main characteristics of this inhomogeneous condensate and derive the critical value of the magnetic field. {In Sec. \ref{sec-5} we show how these hadronic configurations behave in the presence of test quarks.} Sec. \ref{sec-6} is devoted to the conclusions.

\section{Preliminaries} \label{sec-2}

The gauged generalized Skyrme model is described by the action
\begin{gather} 
I(U, A_\mu)\ =\ \frac{f_\pi^2}{4}\int_{\mathcal{M}} d^{4}x \sqrt{-g} \biggl[\text{Tr}\left( R_{\mu }R^{\mu }+\frac{%
\lambda_{Sk} }{8}G_{\mu \nu }G^{\mu \nu }\right) -m_\pi^2 \text{Tr}(2 \mathbb{I} -U-U^{-1})  \biggl]  
-\frac{1}{4}\int d^{4}x\sqrt{-g} F_{\mu \nu } F^{\mu
\nu } + I_{\text{corr}} \ , 
\notag \\ 
R_{\mu }=U^{-1}D_{\mu }U
\ ,\qquad  \qquad G_{\mu \nu }=[R_{\mu},R_{\nu }]\ , \qquad \qquad  F_{\mu \nu }=\partial _{\mu }A_{\nu }-\partial_{\nu }A_{\mu } \ ,\label{I} 
\end{gather}
where $U(x)\in SU(2)$ is the Skyrme field, and the covariant derivative associated to the $U(1)$ gauge field $A_{\mu }$ is defined as $D_{\mu }U=\nabla _{\mu }U+A_{\mu } \left[
t_{3},U\right]$, 
where $t_{a}=i\sigma _{a}$ are the $SU(2)$ generators written in terms of the Pauli matrices $\sigma _{a}$. 
Here, the coupling $f_{\pi }$ is the pion decay constant, $m_\pi$ is the pion mass, and $\lambda_{Sk}$ is a positive number fixed experimentally. Moreover, $I_{\text{corr}}$ represents the sub-leading corrections to the Skyrme model that come from the large-$N_c$ expansion of QCD. In this section, we only consider the usual gauged Skyrme model, however, our construction also works when such terms are included; see Appendix A for the details.

The variation of the action in Eq. \eqref{I} with respect to the fields $U$ and $A_{\mu }$ leads to the following field equations
\begin{align}
&D_{\mu }\left( R^{\mu }+\frac{\lambda_{Sk} }{4}[R_{\nu },G^{\mu \nu
}]\right) - \frac{m_\pi^2}{2}(U-U^{-1}) = 0  \ , 
\qquad
&\nabla _{\mu } F^{\mu \nu } =
-\frac{f^2_\pi}{4}\text{Tr}\Bigr[[t_3,U]\Bigr(R^{\nu }+\frac{\lambda_{Sk} }{4}%
[R_{\mu },G^{\nu \mu }]\Bigr)U ^{-1}\Bigr] \ .
\label{EqMax}
\end{align}

The energy-momentum tensor of the gauged Skyrme model is 
\begin{align} \notag
T_{\mu \nu }=& -\frac{f_\pi^2}{2}\text{Tr}\left[ R_{\mu }R_{\nu }-\frac{1%
}{2}g_{\mu \nu }R^{\alpha }R_{\alpha }+\frac{\lambda_{Sk} }{4}\left( g^{\alpha
\beta }G_{\mu \alpha }G_{\nu \beta }-\frac{1}{4}g_{\mu \nu }G_{\sigma \rho
}G^{\sigma \rho }\right) - \frac{m_\pi^2}{2} g_{\mu\nu} (2\mathbb{I}-U-U^{-1}) \right] 
+
\\ &   
+ g^{\alpha \beta }F _{\mu \alpha
} F_{\nu \beta }-\frac{1}{4}g_{\mu \nu }F^{\alpha \beta} F_{\alpha \beta } \ .  \label{TMUNU}
\end{align}
The topological current $\rho^{\mu}$ is written as
\begin{equation} \label{rho}
\rho^{\mu} = 
\epsilon ^{\mu \nu \alpha \beta }\text{Tr}\Bigr[R_{\nu }R_{\alpha}R_{\beta }
-3\nabla _{\nu }[A_{\alpha }t_{3}(U^{-1}\nabla _{\beta
}U+(\nabla _{\beta }U)U^{-1})]
\Bigr] \ , 
\end{equation} 
where the first term is the usual topological 
density, while the second is the so-called Callan-Witten term, which must be added in order to preserve the current conservation and gauge invariance. 
The integral of the temporal component of the topological current on a space-like hypersurface represents the topological (baryonic) charge of the configuration
\begin{equation}
n_{\text{B}}=\frac{1}{24\pi ^{2}}\int_{\Sigma }\rho^{0}, \qquad \qquad \rho^{0}=\epsilon ^{ijk}\text{%
Tr}\Bigr[R_{i}R_{j}R_{k}-3\nabla _{i}[A_{j}t_{3}(U^{-1}\nabla _{k}U+(\nabla
_{k}U)U^{-1})]\Bigr] \ .  \label{B}
\end{equation}
It is also possible to add a baryon chemical potential, $\mu_B$, via the Wess-Zumino-Witten term, given by 
\begin{equation} \label{WZW}
 I_{\text{WZW}} = \frac{1}{24 \pi^2}\int d^4 x \biggl( \mu_B \rho^0- \frac{1}{2} A_\mu \rho^\mu\biggl)   \ . 
\end{equation}
\bigskip

A general element of $SU(2)$ in the exponential representation is written as
\begin{align}
U^{\pm 1}(x^{\mu })=\cos \left( \alpha \right) \mathbf{1}_{2}\pm \sin \left(
\alpha \right) n^{i}t_{i}\ \qquad \mbox{with} \qquad 
\label{hedgehog}
\left\{
\begin{array}{l}
n^{1}=\sin \Theta \cos \Phi \ ,     \\
n^{2}=\sin \Theta \sin \Phi \ ,     \\
n^{3}=\cos \Theta \ ,
\end{array}
\right.
\end{align}
where $\alpha$, $\Theta$ and $\Phi$ are the three degrees of freedom of the $SU(2)$ field.  A relevant fact comes from the Callan-Witten term in this representation. One can check that, by replacing Eq. \eqref{hedgehog} into Eq. \eqref{rho}, both contributions of the topological charge density are expressed in terms of the degrees of freedom $\alpha$, $\Theta$, and $\Phi$, as follows 
\begin{align} 
    \rho^{0}
    = 12 \sin^2\alpha \sin\Theta \, \mathrm{d}\alpha \;\mathrm{d}\Theta \;\mathrm{d} \Phi 
    +
    12 \;\mathrm{d} \left[  \left( \alpha - \frac{1}{2} \sin(2 \alpha) \right) \sin \Theta A \; \mathrm{d} \Theta -  \alpha   \cos\Theta F   \right]    \ \label{Callan} ,
\end{align}
where $A=A_i \mathrm{d} x^i$ and $F= \frac{1}{2} F_{ij} \mathrm{d}x^i \mathrm{d}x^j$.
It is evident that fixing any one of the degrees of freedom to a constant implies that the contribution of the usual topological charge vanishes. However, the Callan-Witten term 
is not necessarily zero
under this assumption. For example, fixing $\Theta=\pi$,
the topological charge density receives a non-vanishing contribution from the Callan-Witten term, reading
\begin{equation}\label{qlo}
 \rho^{0}=
 -12
 F \, \mathrm{d} \alpha \ .
\end{equation}
Therefore, one can construct topological solitons considering only two non-trivial degrees of freedom, as long as there is a non-zero electromagnetic field strength. A similar strategy can also be used to construct gravitating solutions analytically, with both baryonic charge and magnetic field (see \cite{New1},\cite{New2}, and references therein).

\section{Gauged solitons at finite volume} \label{sec-3}

    Let us consider a finite volume system described by the metric
    \begin{equation} \label{box}
        ds^2= - dt^2 + dx^2 + dy^2 + dz^2 \ ,
    \end{equation}
    where the spatial coordinates $\{x, y, z\}$ have the ranges 
    \begin{equation} \label{ranges}
    -L/2 \leq x \leq L/2 \ ,  \quad 0 \leq y \leq \pi \ , \quad 0 \leq z \leq 2\pi \ .
    \end{equation}
According to the discussion in the previous section, a simple choice for the matter field that considerably reduces the field equations and, at the same time, leads to a non-vanishing topological charge, is
\begin{equation} 
\alpha =\alpha (t, x)\ , \qquad\qquad \Theta =\pi \ ,  \label{ansatzU}
\end{equation}
together with the Maxwell potential that leads to a constant magnetic field in the $x$ direction, that is
\begin{equation} \label{ansatzA}
    A_\mu =  
    (0, 0, B 
    z, 0) \ .
\end{equation}
For this Ansatz, the Maxwell equations are automatically satisfied, while the Skyrme equations are reduced to the sine-Gordon one
\begin{equation} \label{SG}
    \Box \alpha - m_{\pi}^2 \sin(\alpha) = 0 \ ,
\end{equation}
where $\Box  = -\partial_t^2 + \partial_x^2$.
The non-trivial topological charge density in Eq. \eqref{qlo} comes from the Callan-Witten term, $\rho^0 = {12} B \partial_x \alpha$, resulting in the topological charge
\begin{equation} \label{Bfinal}
    n_{\text{B}}= \frac{1}{24 \pi^2} \int 
    12 
    B
    \, \mathrm{d}\alpha \mathrm{d}y \mathrm{d}z  = 
    B
    (\alpha(x_f)- \alpha(x_i)) \ . 
\end{equation}
Assuming the boundary conditions $\alpha(-L/2)=0$ and $\alpha(L/2)=2n \pi$, with $n$ as an integer (a condition that guaranties the periodicity of the $U$ matrix: $U(-L/2)=  U(L/2)$), the topological charge turns out to be
\begin{equation} \label{nB}
n_{\text{B}}= 2n \pi\,B
\ .
\end{equation}
Note that, as the topological charge is the baryon number, which is an integer, $2\pi B$ must also be an integer. While this quantization arises from the boundary conditions in a finite volume, its discrete nature becomes negligible in the high-baryon-density regime. In this limit, a unit change in baryon number represents an infinitesimal fraction of the total charge, allowing both the baryonic charge and the magnetic field to be safely treated as continuous variables.

In the static case, $\alpha=\alpha(x)$, the sine-Gordon equation is reduced to
\begin{equation} \label{Eqa0}
     \frac{\mathrm{d}^2 \alpha}{\mathrm{d} x^2}  - m_\pi^2 \sin(\alpha) = 0 \ . 
\end{equation}
This equation can be rewritten as the following quadrature,
\begin{equation} \label{eq:quadrature}
    \left(\frac{\mathrm{d} \alpha}{\mathrm{d} x} \right)^2  + 2 m^2 _\pi \cos \alpha = \ell \ ,
\end{equation}
 where $\ell$ is an integration constant with dimensions of energy squared. A general solution for this equation is given by the Jacobi amplitude Am$(x,m)$, as
\begin{equation}
\alpha(x) =    2 \mathrm{Am}\!\left(\frac12\sqrt{\ell-2m_\pi^2}\left(x+\frac L2\right),\, \frac{4 m^2_\pi}{ 2m^2_\pi - \ell}\right) \ .
\end{equation}

The resulting baryonic density $\rho^0$ reads
\begin{equation}
 \rho^0 =
12 B \sqrt{\ell - 2m^2_\pi} \; \mathrm{dn}\left(\frac12\sqrt{\ell-2m_\pi^2}\left(x+\frac L2\right),\, \frac{4 m^2_\pi}{ 2m^2_\pi - \ell}\right) \ ,
\end{equation}
where $\dn(x,m)$ is the Jacobi elliptic function. It is periodic in $x$ with  a period given by the complete elliptic integral of the first kind $K(m)$ as $4K\left[4 m^2_\pi/(2m^2_\pi - \ell)\right]/\sqrt{\ell-2m_\pi^2}$. This period is real while $\ell>2m_\pi^2$, being divergent at $\ell=2m_\pi^2$, and approaches $2\pi$ as $\ell \to \infty$, which means increasingly shorter $K(m)$ periods for the baryonic density.

Finally, since $|x| < L/2$, we see that the integration constant can be related to both the boundary conditions and the baryonic charge by imposing that $\rho^0$ must be $n-$periodic inside the interval, that is 

\begin{equation} \label{pell}
    P(\ell) \equiv \, \frac{4\,K\!\left(\frac{4 m^2_\pi}{2m^2_\pi - \ell}\right)}{\sqrt{\ell - 2 m^2_\pi}} = \frac{L}{n} \ . 
\end{equation}
This equation has a finite number of solutions for $\ell$ as long as $L>2\pi$. 

Here, it is convenient to define the following dimensionless quantities 
\begin{eqnarray}\label{rescaling1}
\tilde \ell=\ell/m_\pi^2 \ ,\qquad \tilde L= L \ m_\pi \ ,  
\end{eqnarray}
such that Eq. \eqref{pell} can be rewritten as
\begin{equation} \label{adimpell}
    m_\pi P(\tilde \ell) \equiv \, \frac{4\,K\!\left(\frac{4 }{2 - \tilde \ell}\right)}{\sqrt{\tilde\ell - 2}} = \frac{\tilde L}{n} \ . 
\end{equation}

In Fig. \ref{figu} we have represented the solutions of Eq. \eqref{adimpell}, which intersect the blue curve ${K\left(\frac{4 }{2 - \tilde \ell}\right)}/{ \sqrt{\tilde \ell - 2}}$. As a result of this periodic matching, the baryonic density is a comb-like function with $n$ lumps inside the box of large $L$, as depicted in Fig. \ref{figrho}.
\begin{figure}
    {
    \includegraphics[width=0.45\linewidth]{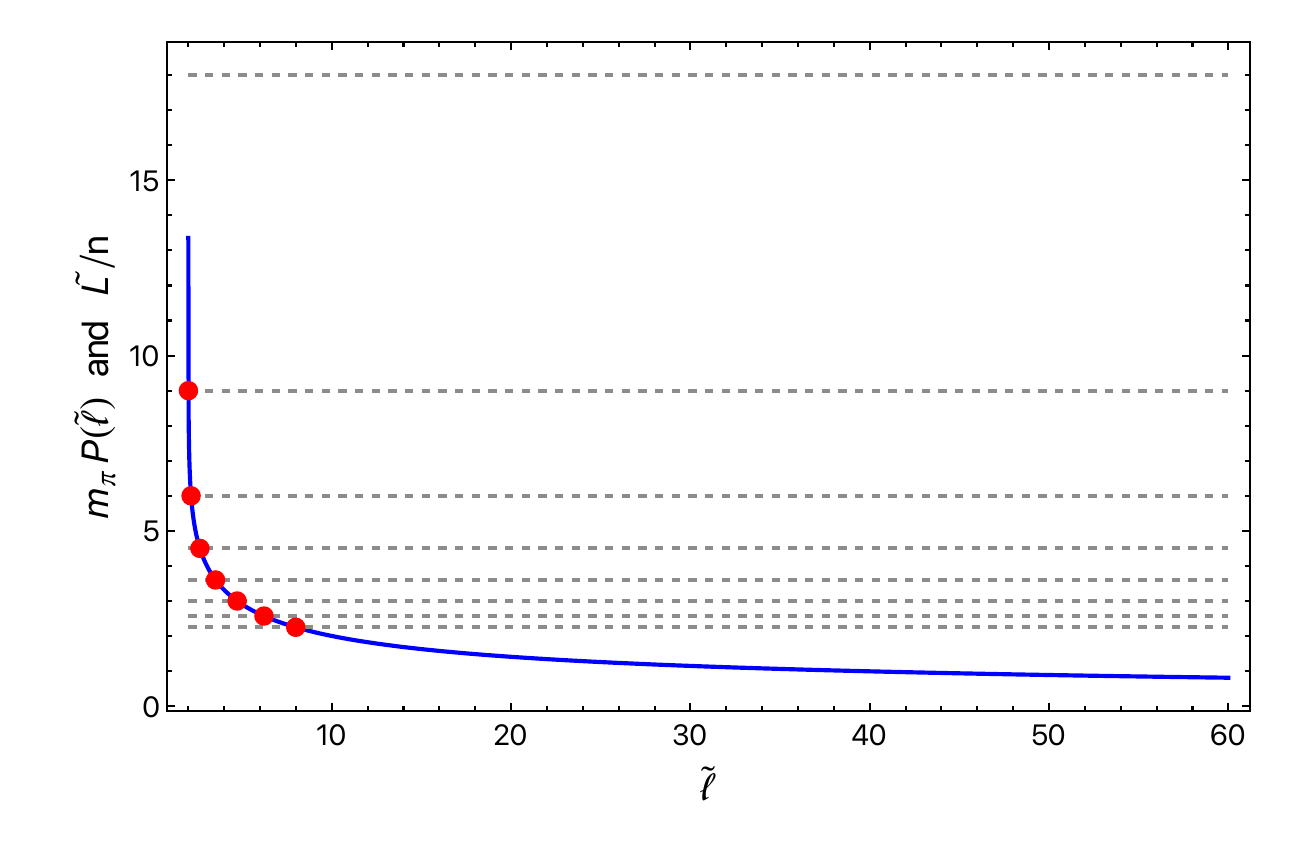} 
    }
    \caption{Numerical solutions of Eq. \eqref{adimpell} illustrating the dependence on $\tilde\ell$. Horizontal dashed lines correspond to $\tilde L/n$, with $\tilde L = 18$. The integer $n$ goes from $1$ to $5$.} 
    \label{figu}
\end{figure}
\begin{figure}
    {
    \includegraphics[width=0.6\linewidth]{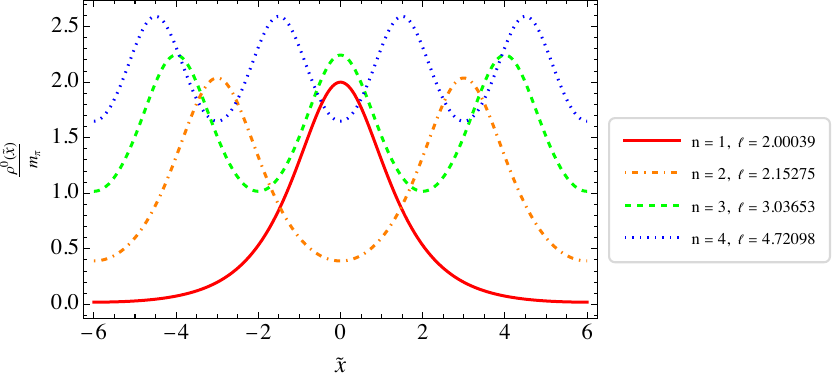}
    }
    \caption{Baryonic density (in units of $m_\pi$) as a function of $\tilde x=x m_\pi$ obtained from Eq. \eqref{rho}, with $\tilde L=18$ and $B=1$.} 
    \label{figrho}
\end{figure}

\section{Emergent chiral soliton lattice} \label{sec-4}

The field configuration presented in the previous section, through the description of topological solitons in a finite volume in the context of the generalized Skyrme theory, is nothing other than (a generalization of) the ChSL. 
To see this more clearly, we can calculate the Hamiltonian density from our Ansatz, obtaining the following
\begin{equation} \label{H}
    \mathcal{H} = f_\pi^2\biggl( \frac{1}{2}(\alpha')^2 +m_\pi^2(1-\cos(\alpha))\biggl) - \frac{\mu_B \, B}{4\pi^2} \alpha' + \frac{B^2}{2} \ .
\end{equation}
From the above expression, we can see that, in the absence of the last term (which is proportional to $B^2$) the Hamiltonian corresponds to the free energy density of the ChSL; this is the sine-Gordon theory with the coupling between the baryon chemical potential, the magnetic field, and the gradient of the pion field. This configuration describes a periodic array of topological solitons carrying baryon charge and magnetic moment, and breaking parity symmetry \cite{Brauner:2016pko}, \cite{Higaki}.

Using the elementary properties of the Jacobi elliptic functions, one can write the energy in terms of the elliptic modulus (see \cite{Brauner:2016pko} and references therein). Then, by minimizing the energy, it is possible to derive the critical value for the magnetic field above which the ChSL is formed; $B> B_{\text{ChSL}}$. This well-known result is given by
\begin{equation}
    B_{\text{ChSL}} = \frac{16 \pi m_\pi f_\pi^2}{\mu_B} \ . 
\end{equation}
Now, in the general case when ${B}^2$ is different from zero (see Eq. \eqref{H}), the solution can be interpreted as a periodic array of hadronic layers that generates its own magnetic field due to the minimal coupling of the Skyrme model with the Maxwell potential. This configuration is, of course, much more difficult to analyze, however, its chiral limit can be studied. In fact, we see that when $m_\pi \rightarrow 0$, the configuration minimizing the free energy  satisfies
\begin{equation} \label{eq:alphaconstant}
    (\alpha') = \frac{\mu_B B}{4 \pi^2 f_\pi^2} \ .
\end{equation}
Notice that the right-hand side of the above equation provides an estimate of the integration constant $\ell$ in Eq. \eqref{eq:quadrature}. In this case, the free energy density of the system becomes 
\begin{eqnarray}
      \lim_{m_\pi \rightarrow 0}\mathcal{H} = -\frac{1}{2} \biggl( \frac{\mu_B B}{4\pi^2 f_\pi}  \biggl)^2 + \frac{B^2}{2} \ . 
\end{eqnarray}
The chiral limit of the system that includes the term proportional to ${B}^2$ in Eq. \eqref{H} was explored in Ref. \cite{Schmitt1}, showing the similarity of this system with the behavior of ordinary type-II superconductors.

Now, it is important to note that the above results can also be derived by considering the partition function of the system,
$Z = \sum_{n}  \text{exp}\{ -\beta(E_{\text{Cl}}- \mu_B n_B)  \} $. Here, the sum runs over the integer number $n$ in the boundary conditions; $E_{\text{Cl}}$ is the classical energy density in Eq. \eqref{H} that does not include the contribution of the WZW term. In  order to have a convergent sum, 
the baryon chemical potential must satisfy $\mu_B < E_{\text{Cl}}/n_B$. Finally, the dispersion relation of pions in this setup can also be derived in a straightforward way by following  Ref. \cite{Takayama}.

{\section{Propagation of probe quarks} \label{sec-5}}

In the previous section, we have shown that the ChSL emerges naturally from the low-energy limit of QCD by considering topological solitons in flat space-time described by a matter field, as given in Eqs. \eqref{hedgehog} and \eqref{ansatzU}. This framework allows for the study of the ChSL in the presence of test two-flavor quarks via the usual Yukawa-like interaction. While the interaction between Skyrmions and fermions has been widely explored  \cite{balachandran1}--\cite{Niemi}  (see also Ref. \cite{Shnir} for a modern overview), exact solutions of the Dirac equation coupled to Skyrme fields are rare, owing to the major mathematical difficulties involved (see, for instance, Refs. \cite{Krusch1}--\cite{zhao}). Fortunately, we can help to bridge this gap using the ChSL as a skyrmionic background and solve analytically the resulting Dirac equation in the high-density limit.

In what follows, we consider an interaction term in the Lagrangian of the form:
\begin{equation} \label{eq:Lint}
    \mathcal{L}_{\text{int}} = \bar{\Psi} \left( i \gamma^\mu D_\mu {-} g U^{\gamma_5} \right) \Psi \ ,
\end{equation}
where  $\Psi$ denotes a massless spin-isospin spinor; that is, {the probe quarks with $u,d$ flavors}. Here, $D_\mu = (\partial_\mu - i  \hat{Q} A_\mu)$ stands for the standard Maxwell-covariant derivative acting on the space of spinors, and $\hat{Q} = \mathrm{diag}\left\{ 2/3, -1/3\right\}$ represents the charge matrix acting solely on the isospin components.  These test fermions will propagate in the fixed classical ChSL background with no back-reaction of the fermionic sector on the Skyrme-Maxwell configuration, so that the fermionic energy density is not fed back into the energy-momentum tensor nor into the Maxwell source (as previously emphasized, this is a good approximation since the quark mass is much smaller than the soliton mass).

The constant $g$ parametrizes the effective coupling between test quarks and the chiral field. Since $U^{\gamma_5}$ is dimensionless, $g$ carries dimensions of mass and can be interpreted as an effective constituent-fermion mass scale generated by the chiral background, while the corresponding dimensionless pion-fermion coupling is of order $g/f_\pi$ (up to normalization conventions). In effective chiral quark-soliton descriptions, the magnitude of this effective coupling can be estimated from known quark-hadron cross-sections, and it is physically expected to decrease with increasing baryon density. Although a precise calculation of these interaction vertices and their scale evolution remains challenging, realistic bounds for the effective coupling in dense media can be drawn, for instance, from Ref.~\cite{pal1999neutron}.

The matter field is coupled in a minimal, parity-invariant manner, as follows
\begin{equation}
\begin{split}
      U^{\gamma_5} &= \left(\frac{1 +  \gamma_5}{2}\right) U + \left(\frac{1 - \gamma_{5} }{2} \right) U^{-1}  = \cos(\alpha) -  i \gamma_5 \tau_3 \sin(\alpha) \ .
\end{split}
\end{equation}
The Pauli matrices are written as $\tau_a$ to emphasize that they act on the isospin components of the spinor. The resulting Dirac Hamiltonian is given by
\begin{equation}
    \hat{H}  = \vec{\alpha} \cdot \vec{\pi} + g \hat{\beta} U^{\gamma_5} \ ,
\end{equation}
where $\vec{\pi} = \left( \vec{p} -   \hat{Q} \vec{A}  \right)$ is the canonical momentum operator, and $\vec{A} = (0,B z,0)$. Note that all operators involved in the Hamiltonian are diagonal in isospin space; hence, we can treat the Dirac spinor as an eigenstate of $\tau_3$ and separate the system into two decoupled equations, 
\begin{equation}
    \hat{H}_s \Psi^{(s)} = E \Psi^{(s)}  \ ,
\end{equation}
where $s=\pm 1$ denotes the isospin component ($s=1$ for $u$-quarks, $s=-1$ for $d$-quarks). Next, we perform a local chiral rotation on the spinor. Defining
\begin{equation} \label{eq:chiral-rotation}
    \hat{S} = \exp{\left(  \frac{i}{2} \gamma_5 \tau_3 \; \alpha(x) \right) } \ , \qquad  \psi \equiv \hat{S}\Psi \ , 
\end{equation}
so that $\Psi= \hat{S}^{-1} \psi$ and $\bar{\Psi}= \bar{\psi}\hat S^{-1}$, and applying this to the Lagrangian in Eq. \eqref{eq:Lint}, we obtain the following transformed Lagrangian:
\begin{equation} \label{eq:Ltrans}
    \mathcal{L}' = \bar{\psi} \left(i \gamma^\mu D_\mu - g  -\frac 12 \gamma_1  \gamma_5 \tau_3\partial_x \alpha (x) \right) \psi \ .
\end{equation}
The ChSL thus contributes both as a mass term through the coupling constant and as a pseudo-vector potential. This modified system leads to a Hamiltonian where the coupling is manifested explicitly as a constant mass term
\begin{equation}
    \hat{H}_s = \vec{\alpha} \cdot \vec{\pi}_s + \hat{\beta}g + \frac s2 \alpha'(x) \, \alpha^1 \gamma_5 \ . 
\end{equation}
}
This reformulation is convenient as it trades the position-dependent chiral matrix structure for a derivative axial coupling, which is better suited for an analytic treatment. Moreover, this is a standard approach used in the study of Dirac spinors coupled to axionic backgrounds, which feature the same type of chirally-invariant coupling (see, for example, Ref. \cite{smith2024fermionic}).

Owing to the spatial dependence on the $x$ coordinate, boosts and chiral rotations are closely related. Performing a chiral rotation is equivalent to choosing a specific chiral basis for the fermions \cite{Watson}, allowing us to work with the Dirac equation derived from the modified Lagrangian. In general, the system describes a Dirac particle in a constant, uniform magnetic field \cite{Bhattacharya:diracequation} featuring a discrete symmetry along the $x$ direction. As we shall see, exploiting these features allows for the analytical determination of the Dirac spectrum and the corresponding band structure in the high-density limit\footnote{The derivation of the general spectrum beyond this limit remains a challenging task. We hope to return to the general problem with a numerical treatment in a future publication.}. Such a limit plays a fundamental role in the present understanding of neutron stars and heavy-ion collisions (see \cite{Kogut}  and references therein).

\bigskip

\subsection{Decoupling the Dirac equation}
{Here, we will work with the Dirac representation of the gamma matrices,
\begin{equation}
    \gamma^0=
\begin{pmatrix}
\mathbb I&0\\
0&-\mathbb I
\end{pmatrix},
\qquad
\gamma^i=
\begin{pmatrix}
0&\sigma_i\\
-\sigma_i&0
\end{pmatrix},\qquad
\gamma_5=
\begin{pmatrix}
0&\mathbb I\\
\mathbb I&0
\end{pmatrix} \ , 
\end{equation}
so that, 
\begin{equation}
\beta=\gamma^0=
\begin{pmatrix}
\mathbb I&0\\
0&-\mathbb I
\end{pmatrix},
\qquad
\alpha^i=
\begin{pmatrix}
0&\sigma_i\\
\sigma_i&0
\end{pmatrix},    
\end{equation}
and the term containing the $x$-dependence is block-diagonal,
\begin{equation}
   \alpha^1 \gamma_5 =\begin{pmatrix}
\sigma_1&0\\
0&\sigma_1
\end{pmatrix} \ . 
\end{equation}

In order to decouple the Dirac equation, it is convenient to have the above term fully diagonal, so we can instead choose a ``rotated" version of the Dirac representation\footnote{{This choice is made for convenience, such that the diagonal gamma matrix corresponds to the direction of the magnetic field. Note that in the literature, the magnetic field is sometimes chosen along the $z$ direction, so no additional rotation is needed.}}
 \begin{equation}
   \widetilde\sigma_x=\sigma_3 \ ,
\qquad
\widetilde\sigma_y=\sigma_1 \ ,
\qquad
\widetilde\sigma_z=\sigma_2 \ . 
 \end{equation}
Writing $\psi^{(s)}=(\phi_1,\phi_2,\chi_1,\chi_2)^T$, the complete Hamiltonian eigenvalue problem can be written as
\begin{equation}\label{eq:fullHs-matrix}
H_s\psi^{(s)}=E\psi^{(s)},\qquad
H_s=
\begin{pmatrix}
 g+\frac{s}{2}\alpha'&0&-i\partial_x&\mathcal D_{s,-}\\
 0&g-\frac{s}{2}\alpha'&\mathcal D_{s,+}&i\partial_x\\
 -i\partial_x&\mathcal D_{s,-}&-g+\frac{s}{2}\alpha'&0\\
 \mathcal D_{s,+}&i\partial_x&0&-g-\frac{s}{2}\alpha'
\end{pmatrix},
\end{equation}
where we have defined the Landau-level operator,
\begin{equation}
\mathcal D_{s,\mp}\equiv p_y-\omega_s z\mp\partial_z \ .
\end{equation}
Here, $\omega_s =  Q_s B$, where $Q_s$ is the corresponding entry of the charge matrices ($Q_1 = 2/3, \, Q_{-1} = -1/3$). 

Equivalently, Eq.~\eqref{eq:fullHs-matrix} gives the four-component equations 
\begin{align} 
\left(E-g-\frac{s}{2}\alpha'\right)\phi_1
 &= -i\partial_x\chi_1+\mathcal D_{s,-}\chi_2  \ , \label{ham41}\\
\left(E-g+\frac{s}{2}\alpha'\right)\phi_2
 &= +i\partial_x\chi_2+\mathcal D_{s,+}\chi_1 \ , \label{ham42}\\
\left(E+g-\frac{s}{2}\alpha'\right)\chi_1
 &= -i\partial_x\phi_1+\mathcal D_{s,-}\phi_2 \ , \label{ham43}\\
\left(E+g+\frac{s}{2}\alpha'\right)\chi_2
 &= +i\partial_x\phi_2+\mathcal D_{s,+}\phi_1 \ . \label{ham44}
\end{align}
In this system, the Landau magnetic operators appear explicitly for each component.
For each charge sector $s$, let us define
\begin{equation}\label{eq:guiding-centers}
 z_{0,s}=\frac{p_y}{\omega_s},\qquad
 \xi_s=\sqrt{|\omega_s|}\,(z-z_{0,s}) \ ,
\end{equation}
and the usual Landau-level functions, given by the Hermite polynomials,
\begin{equation}
\varphi_\lambda^{(s)}(z-z_{0,s})=
\left(\frac{\sqrt{|\omega_s|}}{\pi^{1/2}2^\lambda \lambda!}\right)^{1/2}
H_\lambda(\xi_s)e^{-\xi_s^2/2} \ ,
\qquad \varphi_{-1}^{(s)}\equiv0 \ ,
\end{equation}
with $\lambda$ labeling each level. We next analyze each sector independently, demonstrating how an appropriate ansatz decouples the $x$ and $z$ components of the equation.

\subsubsection{The $s=+1$ sector}

In the case when $s=+1$, one has $\omega_+>0$. Being
\begin{equation}
 a_+=\frac{1}{\sqrt2}(\xi_++\partial_{\xi_+}) \ ,\qquad
 a_+^\dagger=\frac{1}{\sqrt2}(\xi_+-\partial_{\xi_+}) \ ,
\end{equation}
the level operators are written as 
\begin{equation}\label{eq:landau-plus-ops}
\mathcal D_{+,-}=-\sqrt{2|\omega_+|}\,a_+ \ ,
\qquad
\mathcal D_{+,+}=-\sqrt{2|\omega_+|}\,a_+^\dagger \ .
\end{equation}
For an excited level $\lambda\geq1$, a suitable Ansatz for the spinor components is the following
\begin{equation}\label{eq:ansatz-plus}
\phi_1=f_{+1}(x)\varphi_{\lambda-1}^{(+)} \ ,\qquad
\phi_2=f_{+2}(x)\varphi_{\lambda}^{(+)} \ ,\qquad
\chi_1=h_{+1}(x)\varphi_{\lambda-1}^{(+)} \ ,\qquad
\chi_2=h_{+2}(x)\varphi_{\lambda}^{(+)} \ .
\end{equation}
Using Eq.~\eqref{eq:landau-plus-ops}, the problem reduces to a system of coupled ordinary differential equations in the $x$ variable, namely, 

\begin{align}\label{eq:plus-odes}
\left(E-g-\frac{\alpha'}2\right)f_{+1}
 &= -ih_{+1}'-\Omega_{+,\lambda}h_{+2} \ ,\\
\left(E-g+\frac{\alpha'}2\right)f_{+2}
 &= +ih_{+2}'-\Omega_{+,\lambda}h_{+1} \ ,\\
\left(E+g-\frac{\alpha'}2\right)h_{+1}
 &= -if_{+1}'-\Omega_{+,\lambda}f_{+2} \ ,\\
\left(E+g+\frac{\alpha'}2\right)h_{+2}
 &= +if_{+2}'-\Omega_{+,\lambda}f_{+1} \ ,
\end{align}
with a separation constant defined as 
\begin{equation}\label{Omplus}
\Omega_{+,\lambda}\equiv\sqrt{2|\omega_+|\lambda} \ ,\qquad \lambda\geq1 \ .
\end{equation}
At $\lambda=0$, Eq.~\eqref{eq:ansatz-plus} contains only the second and fourth components,
\begin{equation}\label{eq:LLL-plus-spinor}
\psi_{+,0}=\begin{pmatrix}
0\\ f_{+}(x)\varphi_0^{(+)}\\0\\h_{+}(x)\varphi_0^{(+)}
\end{pmatrix}.
\end{equation}

\subsubsection{The $s=-1$ sector}

For $s=-1$, one has $\omega_-<0$. In this case, the roles of the raising and lowering operators are interchanged:
\begin{equation}\label{eq:landau-minus-ops}
\mathcal D_{-,-}=+\sqrt{2|\omega_-|}\,a_-^\dagger \ ,
\qquad
\mathcal D_{-,+}=+\sqrt{2|\omega_-|}\,a_- \ .
\end{equation}
Consequently, for $\lambda\geq1$, we should consider the Ansatz:
\begin{equation}\label{eq:ansatz-minus}
\phi_1=f_{-1}(x)\varphi_{\lambda}^{(-)} \ ,\qquad
\phi_2=f_{-2}(x)\varphi_{\lambda-1}^{(-)} \ ,\qquad
\chi_1=h_{-1}(x)\varphi_{\lambda}^{(-)} \ ,\qquad
\chi_2=h_{-2}(x)\varphi_{\lambda-1}^{(-)} \ .
\end{equation}
Once again, the $x$ and $z$ components are decoupled, yielding the component equations:
\begin{align}\label{eq:minus-odes}
\left(E-g+\frac{\alpha'}2\right)f_{-1}
 &= -ih_{-1}'+\Omega_{-,\lambda}h_{-2} \ ,\\
\left(E-g-\frac{\alpha'}2\right)f_{-2}
 &= +ih_{-2}'+\Omega_{-,\lambda}h_{-1} \ ,\\
\left(E+g+\frac{\alpha'}2\right)h_{-1}
 &= -if_{-1}'+\Omega_{-,\lambda}f_{-2} \ ,\\
\left(E+g-\frac{\alpha'}2\right)h_{-2}
 &= +if_{-2}'+\Omega_{-,\lambda}f_{-1} \ ,
\end{align}
with
\begin{equation}\label{Omminus}
\Omega_{-,\lambda}\equiv\sqrt{2|\omega_-|\lambda} \ ,\qquad \lambda\geq1 \ .
\end{equation}
In this charge sector, we observe that at order $\lambda=0$, only the first and third components survive, that is
\begin{equation}\label{eq:LLL-minus-spinor}
\psi_{-,0}=\begin{pmatrix}
f_{-}(x)\varphi_0^{(-)}\\0\\h_{-}(x)\varphi_0^{(-)}\\0
\end{pmatrix} \ .
\end{equation}

\subsection{The high-density limit}

As previously noted, in the high baryonic density limit, the profile $\alpha(x)$ exhibits linear behavior, taking the form $\alpha(x) = ax + \alpha_0$. From Eq.  \eqref{nB}, we find that $a = n_B/B L$ and $\alpha_0 = a L/2$, which will be used later in the numerical analysis. In this regime, the Dirac equation can be solved analytically. Let us consider the following plane-wave Ansatz for the $\lambda\geq1$ Dirac spinors:
\begin{equation}
f_{sj}(x) = F_{sj} e^{ikx} \ , \qquad h_{sj}(x) = H_{sj} e^{ikx} \ , \qquad j=1,2.
\end{equation}
With this, we will now derive the spectrum independently in the two charge sectors directly from the complete $4\times4$ Hamiltonian blocks.

\subsubsection{The $s=+1$ sector}

First, substituting the plane-wave Ansatz into Eq.~\eqref{eq:plus-odes} gives
\begin{align}
\left(E-g-\frac a2\right)F_{+1}&=kH_{+1}-\Omega_{+,\lambda}H_{+2} \ ,\\
\left(E-g+\frac a2\right)F_{+2}&=-kH_{+2}-\Omega_{+,\lambda}H_{+1} \ ,\\
\left(E+g-\frac a2\right)H_{+1}&=kF_{+1}-\Omega_{+,\lambda}F_{+2} \ ,\\
\left(E+g+\frac a2\right)H_{+2}&=-kF_{+2}-\Omega_{+,\lambda}F_{+1} \ .
\end{align}
Therefore, we obtain
\begin{equation}\label{eq:Hplus-block}
H_{+,\lambda}=
\begin{pmatrix}
 g+\frac a2&0&k&-\Omega_{+,\lambda}\\
 0&g-\frac a2&-\Omega_{+,\lambda}&-k\\
 k&-\Omega_{+,\lambda}&-g+\frac a2&0\\
 -\Omega_{+,\lambda}&-k&0&-g-\frac a2
\end{pmatrix},\qquad \lambda\geq1 \ .
\end{equation}
Now, the characteristic polynomial factorizes as
\begin{equation}\label{eq:charpoly-plus}
\begin{split}
P_{+,\lambda}(E)&\equiv\det\!\left(E\mathbb I_4-H_{+,\lambda}\right)\\
&=\left[E^2-\Omega_{+,\lambda}^2-\left(\sqrt{g^2+k^2}+\frac a2\right)^2\right]
\left[E^2-\Omega_{+,\lambda}^2-\left(\sqrt{g^2+k^2}-\frac a2\right)^2\right] \ ,
\end{split}
\end{equation}
so that, there are four branches in this sector
\begin{equation}\label{eq:spectrum-plus}
E_{+,\lambda,\eta,\zeta}(k)=
\eta\sqrt{\Omega_{+,\lambda}^2+
\left(\sqrt{g^2+k^2}+\zeta\frac a2\right)^2} \ ,
\qquad \eta,\zeta=\pm 1 \ .
\end{equation}
Here, the $\eta$ factor labels the overall sign of the relativistic energy and hence it naturally identifies particle or anti-particle states, while $\zeta$ is related to spin-polarization in the chiral background.

\subsubsection{The $s=-1$ sector}

Following the same steps as the previous subsection, for $s=-1$, Eq.~\eqref{eq:minus-odes} gives
\begin{align}
\left(E-g+\frac a2\right)F_{-1}&=kH_{-1}+\Omega_{-,\lambda}H_{-2} \ ,\\
\left(E-g-\frac a2\right)F_{-2}&=-kH_{-2}+\Omega_{-,\lambda}H_{-1} \ ,\\
\left(E+g+\frac a2\right)H_{-1}&=kF_{-1}+\Omega_{-,\lambda}F_{-2} \ ,\\
\left(E+g-\frac a2\right)H_{-2}&=-kF_{-2}+\Omega_{-,\lambda}F_{-1} \ .
\end{align}
It follows that
\begin{equation}\label{eq:Hminus-block}
H_{-,\lambda}=
\begin{pmatrix}
 g-\frac a2&0&k&+\Omega_{-,\lambda}\\
 0&g+\frac a2&+\Omega_{-,\lambda}&-k\\
 k&+\Omega_{-,\lambda}&-g-\frac a2&0\\
 +\Omega_{-,\lambda}&-k&0&-g+\frac a2
\end{pmatrix},\qquad \lambda\geq1 \ .
\end{equation}
In this sector, the characteristic polynomial takes the form
\begin{equation}\label{eq:charpoly-minus}
\begin{split}
P_{-,\lambda}(E)&\equiv\det\!\left(E\mathbb I_4-H_{-,\lambda}\right)\\
&=\left[E^2-\Omega_{-,\lambda}^2-\left(\sqrt{g^2+k^2}+\frac a2\right)^2\right]
\left[E^2-\Omega_{-,\lambda}^2-\left(\sqrt{g^2+k^2}-\frac a2\right)^2\right] \ .
\end{split}
\end{equation}
Therefore
\begin{equation}\label{eq:spectrum-minus}
E_{-,\lambda,\eta,\zeta}(k)=
\eta\sqrt{\Omega_{-,\lambda}^2+
\left(\sqrt{g^2+k^2}+\zeta\frac a2\right)^2},
\qquad \eta,\zeta=\pm1.
\end{equation}
Notice that the two charge sectors have the same branch structure but scale differently since $|Q_+| \neq |Q_-|$. In general, we can write,
\begin{equation}\label{eq:spectrum-unified-n}
E_{s,\lambda,\eta,\zeta}(k)=
\eta\sqrt{\Omega_{s,\lambda}^2+
\left(\sqrt{g^2+k^2}+\zeta\frac a2\right)^2} \ ,
\qquad \lambda\geq1 \ ,
\end{equation}
where $\Omega_{s,\lambda}=\sqrt{2|Q_sB|\lambda}$.

\subsubsection{Lowest Landau Level}

The LLL must be treated separately. From Eqs.~\eqref{ham41}-\eqref{ham44} and the spinors in Eqs.~\eqref{eq:LLL-plus-spinor} and \eqref{eq:LLL-minus-spinor}, we can see that both charge sectors reduce to an effective 1-dimensional Dirac equation problem (actually, this dimensional reduction holds even out of the linear limit at high baryonic density). Their respective Hamiltonians  are, for $s=+1$,
\begin{equation} \label{eq:hamLLLplus}
H_{+,0}=\begin{pmatrix}g-\frac a2&-k\\-k&-g-\frac a2\end{pmatrix},
\end{equation}
while for $s=-1$,
\begin{equation} \label{eq:hamLLLminus}
H_{-,0}=\begin{pmatrix}g-\frac a2&+k\\+k&-g-\frac a2\end{pmatrix}.
\end{equation}
From the above, the following dispersion relation is obtained
\begin{equation}\label{eq:LLL-spectrum-new(k)}
E_{s,0,\eta}(k)=-\frac a2+\eta\sqrt{g^2+k^2} \  ,\qquad \eta=\pm1 \ .
\end{equation}
We emphasize that the results obtained thus far correspond to the chirally rotated spinor defined by the transformation in Eq.~\eqref{eq:chiral-rotation}.
The original spinor is, instead,
\begin{equation}\label{spinor-original}
    \Psi_s(x) = \hat{S}^{-1}_s  \psi_s = e^{-s \frac{i}{2} \gamma_5 \alpha(x)} e^{ikx} \Xi_{s,\lambda} \ ,
\end{equation}
and the \textit{original Hamiltonian} (before the chiral rotation) possesses a spatial periodicity $H(x + P_H) = H(x)$. In the high-density limit, the potential goes with $e^{-i s \gamma_5 (a x + \alpha_0))}$, and therefore it has the fundamental period,
\begin{equation} \label{def:ph}
    P_H = \frac{2 \pi}{a}.
\end{equation}
Hence, we impose the Bloch condition (in the \textit{original spinor})
\begin{equation} \label{eq:Bloch-condition}
   \Psi_{s,q}(x + P_H) = e^{iq P_H} \Psi_{s,q}(x) \ ,
\end{equation}
where $q$ represents the Bloch quasi-momentum.

Notice that, however, the chiral rotation itself is not periodic over a period of the Hamiltonian. Instead, 
\begin{equation}
    \hat{S}_s (x +P_H) = e^{-i s \gamma_5 \pi} \hat{S}_s(x) = - \hat{S}_s(x) \ , 
\end{equation}
so that
\begin{align} \label{eq:imposingBloch}
    \Psi_{s}(x+P_H) &= \hat{S_s}(x+P_H) \psi(x+P_H) \nonumber \\ 
    &= - \hat{S}_s(x) e^{ik P_H} \psi(x) \ .
\end{align}
Comparing with Eq. \eqref{eq:Bloch-condition}, it follows that $e^{i q P_H} = -e^{ik P_H}$, hence:
\begin{equation}
    q= k + \frac{\pi}{P_H} = k + \frac a2 \quad \left(\mathrm{mod} \frac{2\pi}{P_H} \right) \ .
\end{equation}
For $x \in [-L/2, L/2]$, we impose periodic boundary conditions; $\Psi_s(x+L) = \Psi_s(x)$, to ensure consistency with the ChSL background in a finite volume. This requires the box length $L$ to contain an integer number $n$ of Hamiltonian cells
\begin{equation}
    L = n P_H \ , \qquad n \in \mathbb{N} \ ,
\end{equation}
where $n$ is just the winding defined  through the boundary condition $\Delta \alpha = 2 n \pi$. Indeed, from Eq.~\eqref{def:ph}, and since $a = \Delta \alpha /L,$ we can see that,
$$n = \frac{\Delta \alpha}{2 \pi} = \frac{L}{P_H}.$$
In the high-density limit, we have the condition $a L = 2 n \pi$. 
Imposing Eq. \eqref{eq:Bloch-condition} through $n$ cells, we find that
\begin{equation}
    e^{iq L} = 1 \quad \Rightarrow \quad q_\nu = \frac{2 \pi \nu}{L} \ ,
\end{equation}
and, in turn,  the momentum $k$ is quantized as
\begin{equation}
    k_\nu \equiv \frac{2 \pi \nu}{L} - \frac{a}{2}  \ .
\end{equation}

\subsection{Energy spectrum}

Through the preceding analysis, we have derived the analytical expressions for the energy spectrum. In terms of the quantized momentum $k_\nu = \frac{2\pi\nu}{L} - \frac{a}{2}$, these can be written as:
\begin{equation}\label{spectrum1}
    E_{s,\lambda, \eta,\zeta}^{(\nu)} = \eta \sqrt{\Omega_{s,\lambda}^2 + \left( {\zeta \frac a2} +  \sqrt{g^2 +k_\nu^2}\right)^2} \ ,
\end{equation}
whereas the LLL spectrum becomes
\begin{equation}\label{eq:LLL-spectrum-new}
E^  {(\nu)}_{s,0,\eta}=-\frac a2+\eta\sqrt{g^2+ k_\nu^2},\qquad \eta=\pm1 \ .
\end{equation}
Two primary observations are worth emphasizing. First, the LLL dispersion relation closely resembles that of a Dirac fermion in a uniform magnetic field in $3+1$ dimensions (see Ref.~\cite{shovkovy2013magnetic}), but with three fundamental differences. The coupling $g$ acts as an effective mass, thereby opening a mass gap in the spectrum. Furthermore, the momentum is no longer continuous; the periodic boundary conditions quantize it via the integer $\nu$, rendering the spectrum discrete. Nevertheless, the effective dimensional reduction characteristic of the LLL remains manifest, as the dispersion retains its one-dimensional relativistic form:
\[
E \sim \pm \sqrt{g^2 + k_\nu^2} \ .
\]
Finally, the chiral background, parameterized by $a$, does not merely deform the gap; it shifts the entire spectrum. This spectral lifting implies that zero-mode crossings may occur for specific values of this constant.

The second observation is that the spectrum outside the LLL contains four branches, whereas the LLL contains only two. This is not accidental. In the higher Landau level sector ($\lambda > 0$), the energy expression involves two independent binary choices: the overall relativistic sign ($\pm$), corresponding to particle and antiparticle branches, and the internal sign in the term $({a/2} \pm \sqrt{g^2 + k_\nu^2})$. This internal sign reflects the non-trivial interplay between the chiral shift induced by the ChSL background and the standard dynamics of a fermionic system in a uniform magnetic field. Consequently, for each fixed set of quantum numbers $(s, \lambda, \nu)$, one obtains four distinct branches. In contrast, in the LLL, one spinorial/Landau component is projected out, causing the second binary label to disappear and leaving only the two standard relativistic branches. Physically, this is the same mechanism behind the well-known reduction of degeneracy in the LLL for relativistic Landau problems: the lowest level is polarized and thus carries fewer dynamical degrees of freedom than the excited states.

{As in the previous numerical results, and in order to illustrate as clearly as possible the role of the various parameters and how they define the energy spectrum, we will consider once again the dimensionless variables defined in Eq. \eqref{rescaling1}, in addition to the following redefinitions: $g=\tilde{g} m_\pi$, $a=\tilde{a} m_\pi$ and $B=\tilde B m_\pi^2$. Consequently, the spectrum defined in Eq. \eqref{spectrum1} can be written as
\begin{equation}\label{redimspectrum1}
    E_{s,\lambda, \pm \pm}^{(\nu)} = \pm m_\pi \sqrt{\tilde \Omega_{s,\lambda}^2 + \left( \pm\frac{\tilde{a}}{2} +  \sqrt{\tilde{g}^2 + \left(\frac{2\pi \nu}{\tilde{L}} - \frac{\tilde{a}}{2}\right)^2}\right)^2}.
\end{equation}
where $\Omega _{s,\lambda}=\tilde\Omega _{s,\lambda} m_\pi$, according to Eqs. \eqref{Omplus} and \eqref{Omminus}, the fact that $\omega_s =  Q_s B$ and the rescaling previously defined.}

In what follows, we will present and describe several numerical results (Figs. 3-8) which illustrate the qualitative dependence of the spectrum on its parameters.
The distinct roles of $\tilde{g}$ and $\tilde{a}$ are clearly visible in the numerical results. In what follows, we will study the effect of the ChSL amplitude through its relation with the baryon number, making use of the equations defined in the high-density limit. The families of curves $E(\tilde g)$ in Fig.~\ref{fig:E(g)varB}, $E(\tilde a)$ in Fig.~\ref{fig:E(B)varB}, and the discrete spectrum in Fig.~\ref{fig:E(level)varA} demonstrate that while increasing either $\tilde g$ or $\tilde a$ tends to separate the branches, they govern different physical aspects. The adimensional coupling $\tilde g$ enters symmetrically inside the square root, controlling the size of the gap in a manner analogous to a mass term. Conversely, the parameter $\tilde a$ shifts the spectrum globally and displaces the quantized momentum. Thus, varying $\tilde a$ can generate branch lifting (and zero-energy crossings in the case of the LLL spectrum), whereas $\tilde g$ primarily dictates the distance between the positive and negative energy sectors.

\begin{figure}[H]
    \centering
    \includegraphics[width=0.65\linewidth]{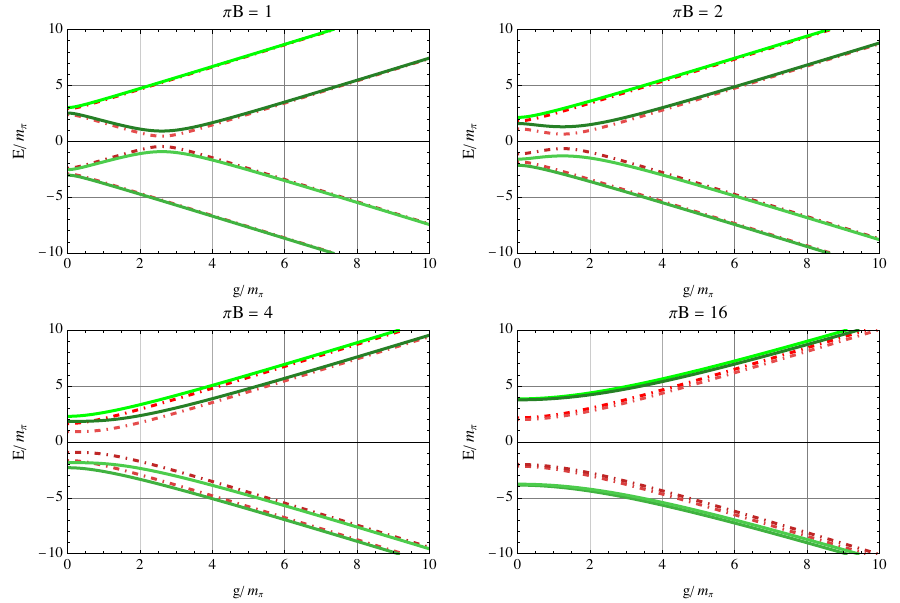}
    \caption{Energy spectrum (in units of $m_\pi$) for $\lambda=1$ as a function of the coupling $\tilde g$, for fixed $\nu=2$ and $n_B =10$, shown for several values of $\tilde {B}$. The parameter $\tilde g$ acts as an effective mass scale, opening a gap, while the magnetic field controls the subsequent branch splitting between isospin states. Isospin $s=1$ states are represented by green solid lines, while $s=-1$ states are shown as red dot-dashed lines.}
    \label{fig:E(g)varB}
\end{figure}

\begin{figure}[H]
    \centering
    \includegraphics[width=0.65\linewidth]{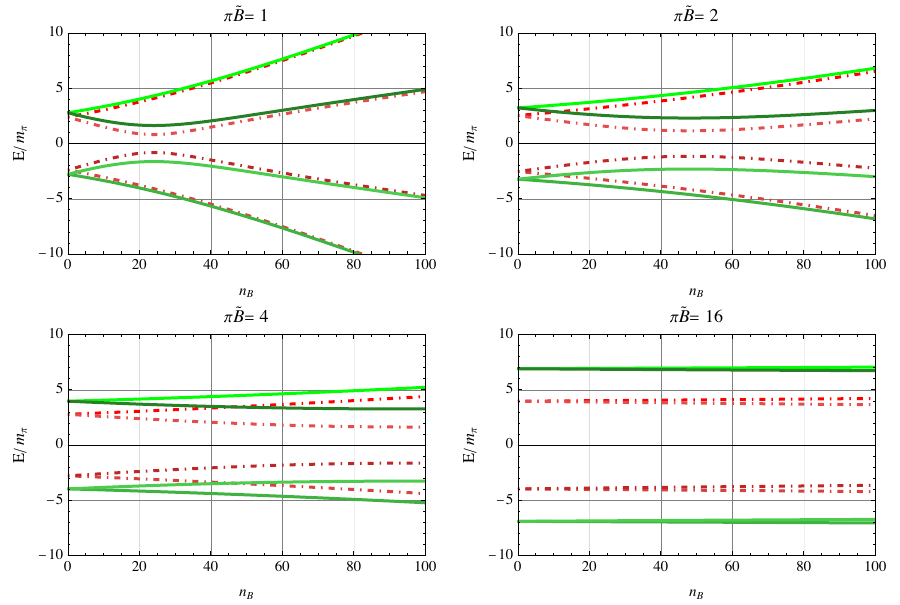}
    \caption{Energy spectrum (in units of $m_\pi$) for $\lambda=1$ as a function of the chiral-background baryonic charge $n_B$, for fixed $\nu=2$ and $\tilde g=2$. The $s=1$ states (green solid lines) and $s=-1$ states (red dot-dashed lines) exhibit asymmetric reorganization. Increasing $\tilde B$ enhances the separation between branches and amplifies the charge-dependent splitting.}
    \label{fig:E(B)varB}
\end{figure}

In this framework, the magnetic field enters only through the separation constant $\tilde\Omega_{s,\lambda}$ and thus affects exclusively the excited Landau levels. This is evident in Fig.~\ref{fig:E(B)varlambda}, where the spectral separation increases with $\tilde B$. Specifically, higher values of $\lambda$ result in larger separations from the low-energy sector, as expected from the growth of the Landau scale. At the same time, the dependence on the isospin label $s$ becomes increasingly pronounced because $\tilde\Omega_{s,\lambda}$ is charge-dependent, and the two isospin components possess charges of different magnitudes. Consequently, the magnetic field splits the positively and negatively charged sectors unequally, as seen in Figs.~\ref{fig:E(level)varA} and \ref{fig:E(B)varlambda}.

\begin{figure}[H]
    \centering
    \includegraphics[width=0.65\linewidth]{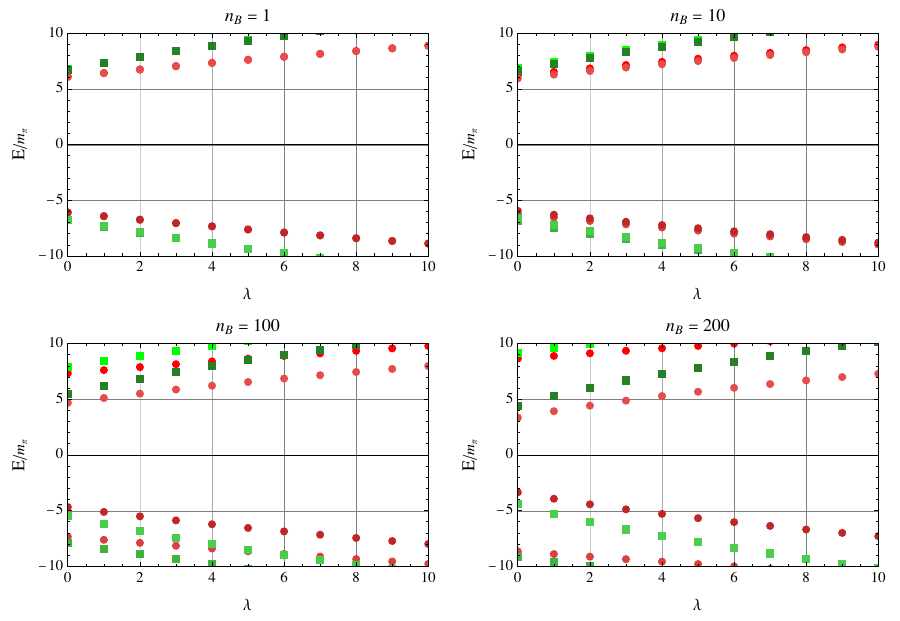}
    \caption{Energy spectrum (in units of $m_\pi$) as a function of $\lambda$, for fixed $\tilde g=4$, $\tilde B=5$, and $\nu=2$, for several values of the baryonic number $n_B$. The baryonic number modifies the spacing between branches. States with $s=1$ are denoted by green squares, and $s=-1$ by red circles.}
    \label{fig:E(level)varA}
\end{figure}
\begin{figure}[H]
    \centering
    \includegraphics[width=0.65\linewidth]{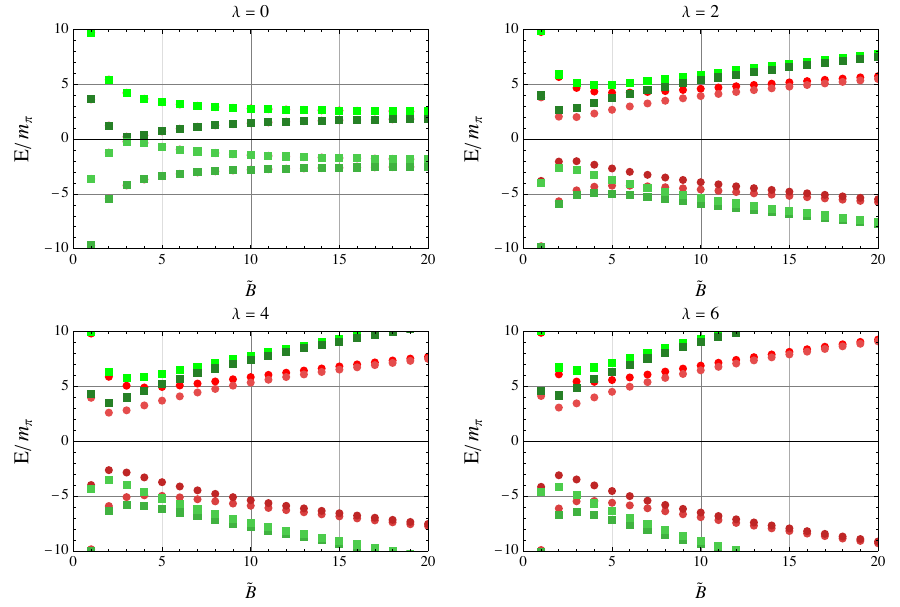}
    \caption{Energy spectrum (in units of $m_\pi$) for different Landau levels $\lambda$ as a function of the magnetic field $\tilde B$, for fixed $\nu=2$, $\tilde g=2$, and $n_B =80$. As the magnetic field increases, the separation between branches of different isospin also increases, and same isospin branches get closer. Green solid lines: $s=1$; red dot-dashed lines: $s=-1$.}
    \label{fig:E(B)varlambda}
\end{figure}

The dependence on the discrete momentum label $\nu$ is illustrated in Fig.~\ref{fig:E(nu)varB} and Fig.~\ref{fig:E(nu)varBnb10000}. Since the spectrum depends on $\nu$ only through the shifted combination $k_\nu = 2\pi\nu/L - a/2$, the minima are not centered at $\nu=0$. This is very visible at high $a$ for large baryonic density and low magnetic field, as shown in Fig.~\ref{fig:E(nu)varBnb10000}.

A comparison between Fig.~\ref{fig:E(nu)varB} and the previous results suggests that the emergence of these multiple branches is an overall effect of the chiral coupling on the fermionic dynamics.

\begin{figure}[H]
    \centering
    \includegraphics[width=0.67\linewidth]{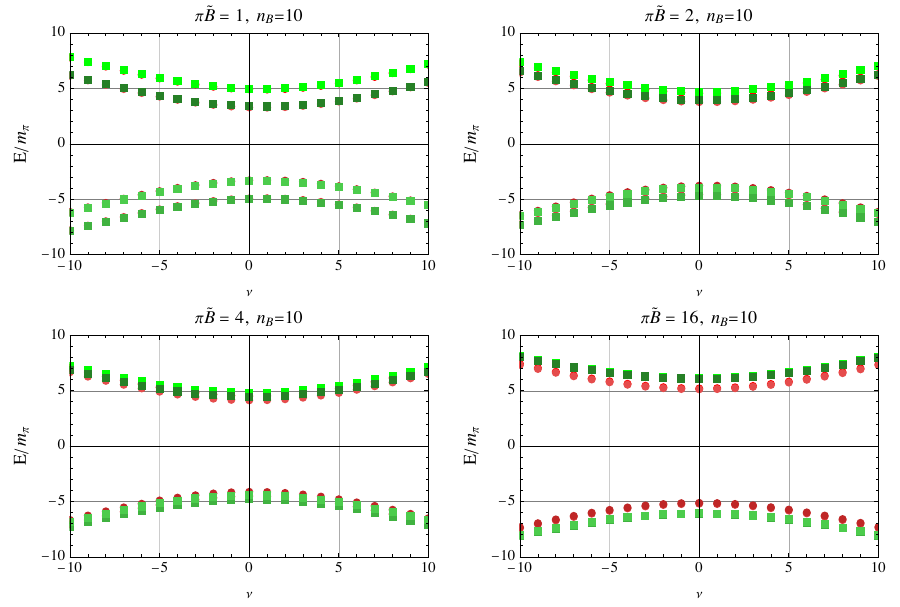}
    \caption{Energy spectrum (in units of $m_\pi$) as a function of the discrete momentum label $\nu$, for fixed $\tilde g=4$, $n_B =10$, and $\lambda=1$. The spectrum is almost symmetrical in $\nu$, and the increase of the magnetic field $B$ tends to separate different isospin branches. Green squares: $s=1$; red circles: $s=-1$.}
    \label{fig:E(nu)varB}
\end{figure}
\begin{figure}[H]
    \centering
    \includegraphics[width=0.67\linewidth]{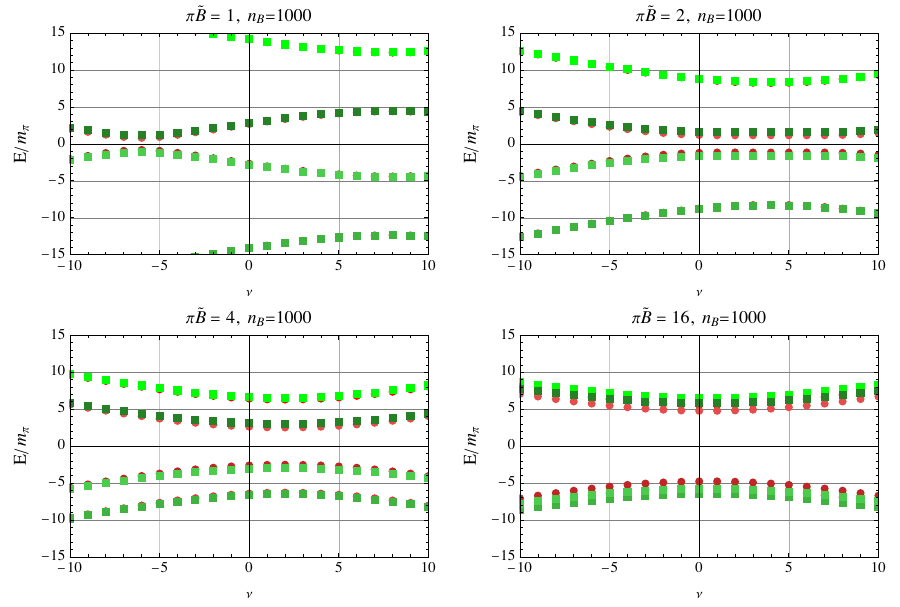}
    \caption{Energy spectrum (in units of $m_\pi$) as a function of the discrete momentum label $\nu$, for fixed $\tilde g=4$ and $\lambda=1$. Here we have considered a significantly higher baryonic charge. At low $\tilde B$ the spectrum is clearly not symmetrical on $\nu$ and the branches are clearly separated in a heavier and lighter sector. As $\tilde B$ increases, these two sectors disappear, the spectrum gets more symmetrical while different charge branches become more separate. Green squares: $s=1$; red circles: $s=-1$.}
    \label{fig:E(nu)varBnb10000}
\end{figure}

\section{Final remarks}  \label{sec-6}

In this paper, we have discussed how the ChSL can be obtained from the Skyrme model minimally coupled with the Maxwell gauge field. We show that our suitable Ansatz fully solves the complete Skyrme-Maxwell equations, rather than treating the electromagnetic field as an external background. Furthermore, a key step has been an Ansatz for the hadronic matter with the property that the only non-vanishing contribution to the baryonic density arises from the Callan-Witten term. Such an Ansatz describes topological solitons at finite baryon density. In this procedure, the fact that the topological charge is an integer (the baryon number in this context) leads directly to a quantization condition for the magnetic field. This quantization is, however, a consequence of the boundary conditions imposed on a finite region. For the ChSL to exist, we require (anti-)periodic boundary conditions, which imply that $\alpha(x)$ evaluates to integer multiples of $\pi$ at the boundaries. In large systems at high baryon density, the discreteness of the magnetic field can be safely neglected.

The main result of the present analysis is that the ChSL configuration remains unchanged when subleading corrections to the Skyrme model in the 't Hooft large-$N_c$ expansion are included. Taking into account the highly non-linear character of such subleading corrections, this is a remarkable result that highlights the robust character of the ChSL. We emphasize that this robustness refers specifically to the subleading corrections within the 't Hooft large-$N_c$ expansion. Conversely, generic higher-derivative operators in the mesonic chiral Lagrangian at next-to-leading order, such as $\mathcal{O}(p^4)$ and $\mathcal{O}(p^6)$ terms involving squared kinetic structures \cite{Gasser:1983yg, Bijnens:1999sh}, could modify the effective sine-Gordon equation of the ChSL phase. A detailed analysis of how these standard chiral corrections alter the soliton profile is left for future work.

Even though the Ansatz in Eq. \eqref{ansatzU} leads to a ChSL configuration, a complete stability proof of the configuration (whether it corresponds to a saddle point or a minimum of the action) is still lacking. To the best of the authors' knowledge, this remains an open problem in the literature. Historically, stability proofs for the Skyrme model have been notoriously difficult to establish, even in the spherically symmetric case with unit baryonic charge (see, e.g., Ref. \cite{MantonBook}). However, two necessary conditions for stability are satisfied in this work: first, the intuitive stability arguments from original references such as Refs. \cite{Brauner:2016pko}--\cite{Brauner:2021sci} still apply to our case; second, small variations in the other chiral angles with $\alpha$ fixed result in an increase in energy. These facts support the expectation that the solution is linearly stable within the Skyrme theory. Nevertheless, when subleading terms are included, the stability analysis becomes significantly more challenging. One crucial conclusion can nonetheless be derived: a necessary condition for the stability of the ChSL is that the energy density must remain positive definite. In other words, it is not sufficient for the coupling constants characterizing the subleading terms to be small; they must also be chosen to ensure the positivity of the energy density.

Furthermore, we have extended this framework by explicitly incorporating the interaction between the ChSL and $u$-$d$ test fermions. By solving the Dirac equation within this inhomogeneous background, we derived the exact analytical spectrum in the high-density limit. This microscopic characterization provides a description of the fermionic excitations, showing how the test quark degrees of freedom are modulated by the periodic structure of the lattice. This result also establishes a clear connection between the effective topological description and the underlying fermionic dynamics of QCD under extreme conditions. 
Notwithstanding this result, there are many natural extensions of our work in which the spectrum may be affected and which are worth exploring. For example, a very interesting one is to consider the inclusion of an anomalous magnetic moment at intense magnetic fields (see \cite{Aguirre} and references therein).
The derivation of the general Dirac spectrum beyond the high-density regime remains an open and challenging problem that we plan to address in a forthcoming work \cite{progress}. 

\acknowledgments

F.C. has been funded by FONDECYT Grant No. 1240048 and by Grant ANID EXPLORACI\'ON
13250014. M. L. has been funded
by FONDECYT Iniciaci\'on No. 11241079. A. V.
has been funded by FONDECYT Iniciaci\'on No. 11261883. N.G. is partially supported by CONICET
grant PIP-2023-11220220100262CO and UNLP grant 2022-11/X931. L.U. has been funded by ANID-Beca de Magíster Nacional 2024-22241713.

\appendix

\section{Generalized Skyrme terms}

In this Appendix, we show that the construction performed above also works when higher-order corrections coming from the large-$N_c$ expansion are supplemented to the Skyrme action. 

The so-called generalized Skyrme model includes the higher-order corrections that come from the large-$N_c$ limit of QCD, which are supplemented in the original Skyrme action. The first two sub-leading terms are given by 
\begin{align*}
\mathcal{L}_{6}=& \frac{c_{6}}{96}\text{Tr}\left[ G_{\mu }{}^{\nu }G_{\nu
}{}^{\rho }G_{\rho }{}^{\mu }\right] \ , \\
\mathcal{L}_{8}=& -\frac{c_{8}}{256}\biggl(\text{Tr}\left[ G_{\mu }{}^{\nu
}G_{\nu }{}^{\rho }G_{\rho }{}^{\sigma }G_{\sigma }{}^{\mu }\right] -\text{Tr%
}\left[ \{G_{\mu }{}^{\nu },G_{\rho }{}^{\sigma }\}G_{\nu }{}^{\rho
}G_{\sigma }{}^{\mu }\right] \biggl)\ .
\end{align*}
Taking into account the above new contributions, the Skyrme equations are given by
\begin{align}
\frac{f_\pi^2}{2}\biggl( D_{\mu }\left( R^{\mu }+\frac{\lambda_{Sk} }{4}[R_{\nu },G^{\mu \nu
}]\right) - \frac{m_\pi^2}{2}(U-U^{-1})  \biggl) +3c_{6}[R_{\mu },D_{\nu }[G^{\rho
\nu },G_{\rho }{}^{\mu }]] &  \notag \\
+4c_{8}\biggl[R_{\mu },D_{\nu }\biggl(G^{\nu \rho }G_{\rho \sigma }G^{\sigma
\mu}+G^{\mu \rho }G_{\rho \sigma }G^{\nu \sigma }+\{G_{\rho \sigma
},\{G^{\mu \rho },G^{\nu \sigma }\}\}\biggl)\biggl] & \ = \ 0\ .
\label{GeneralizedEqs}
\end{align}
The Maxwell equations will also be affected by including these new terms due to the $U(1)$ connection present in each of the corrections. Indeed, the electromagnetic current in the Maxwell equations in Eq. \eqref{EqMax} now takes the form 
\begin{align}
J_{\mu } \ = \ &  - \frac{f_\pi^2}{2}\text{Tr}\biggl[\hat{O}\biggl(R_{\mu }+\frac{%
\lambda_{Sk} }{4}[R^{\nu },G_{\mu \nu }]\biggl)\biggl]+\frac{c_{6}}{32}\text{Tr}%
\biggl[\hat{O}\biggl(\lbrack R^{\alpha },[G_{\mu \nu },G_{\alpha }{}^{\nu }]]%
\biggl)\biggl]  \notag \\
& -\frac{c_{8}}{64}\text{Tr}\biggl[\hat{O}\biggl(\lbrack R^{\alpha
},G_{\alpha }{}^{\nu }G_{\nu }{}^{\rho }G_{\rho \mu }+G_{\rho \mu }G_{\nu
}{}^{\rho }G_{\alpha}{}^{\nu }+\{G^{\nu\rho },\{G_{\mu \nu },G_{\rho\alpha
}\}\}]\biggl)\biggl]\ ,  \label{Jmu}
\end{align}
where we have defined $\hat{O}= U^{-1}[t_3,U]$.
Finally, we can compute the energy-momentum tensor of the gauged generalized Skyrme model, obtaining the following
\begin{equation}  \label{Tmunu}
T_{\mu \nu }^{\text{gen}}\ =T_{\mu\nu} + T_{\mu
\nu }^{(6)}+T_{\mu \nu }^{(8)} \  ,
\end{equation}%
where $T_{\mu\nu}$ has been defined in Eq. \eqref{TMUNU}, and 
\begin{align*}
T_{\mu \nu }^{(6)}=& -\frac{c_{6}}{16}\text{Tr}\biggl(g^{\alpha \gamma
}g^{\beta \rho }G_{\mu \alpha }G_{\nu \beta }G_{\gamma \rho }-\frac{1}{6}%
g_{\mu \nu }G_{\alpha }{}^{\beta }G_{\beta }{}^{\rho }G_{\rho }{}^{\alpha }%
\biggl)\ , \\
T_{\mu \nu }^{(8)}=& \ \frac{c_{8}}{32}\text{Tr}\biggl(g^{\alpha \rho
}g^{\beta \gamma }g^{\delta \lambda }G_{\alpha \mu }G_{\nu \beta }G_{\gamma
\delta }G_{\lambda \rho }+\frac{1}{2}\{G_{\mu \alpha },G_{\lambda \rho
}\}\{G_{\beta \nu },G_{\gamma \delta }\}g^{\alpha \gamma }g^{\beta \rho
}g^{\delta \lambda }  \notag \\
& -\frac{1}{8}g_{\mu \nu }(G_{\alpha }{}^{\beta }G_{\beta }{}^{\rho }G_{\rho
}{}^{\sigma }G_{\sigma }{}^{\alpha }-\{G_{\alpha }{}^{\beta },G_{\rho
}{}^{\sigma }\}G_{\beta }{}^{\rho }G_{\sigma }{}^{\alpha })\biggl)\ .
\end{align*}
At first glance, it looks like a hopeless task to find analytic solutions to the above extremely non-linear field equations. In fact, this is not the case. It is important to note two very relevant facts. The first is that all new contributions to the electromagnetic current are proportional to $\hat{O}$. However, according to the Ansatz in Eqs. \eqref{hedgehog} and \eqref{ansatzU}, the matrix $U$ depends only on the identity matrix and the generator $t_3$, namely,
\begin{equation} \label{newU}
    U(\alpha) = \cos(\alpha)  \mathbf{1}_{2} - \sin(\alpha) t_3 \ . 
\end{equation}
Thus, the tensor $\hat{O}$ is identically zero, and therefore, all extra contributions to the current eventually cancel out. Something similar happens with the field equations in Eq. \eqref{GeneralizedEqs} and the energy density in Eq. \eqref{Tmunu}. Since $U$ is given by Eq. \eqref{newU}, it follows that $R_\mu$ depends only on one of the group's generators. Indeed, in our Ansatz this tensor becomes
\begin{equation}
    R_\mu = - \partial_\mu \alpha \, t_3 \ . 
\end{equation}
From the above, it is clear that the tensor $G_{\mu}=[R_\mu,R_\nu]$ must be $G_{\mu\nu}=0$. Furthermore, since all extra contributions in both the field equations and the energy-momentum tensor depend on this tensor (and not on $R_\mu$ separately), all extra contributions eventually cancel out. From all the above, we can conclude that the solution presented here is not only a solution to the Skyrme model but also to the generalized Skyrme model (similar arguments would also hold, including further terms in the 't Hooft expansion). 

Finally, it is important to mention that one can develop an intuition about how domain-wall-type solitons are unaffected by higher-order terms in the ’t Hooft expansion from the foundational geometric analysis of the Skyrme model in vacuum by Manton \cite{Manton:1987xt}. However, the present work investigates a different physical regime involving finite baryon density and magnetic fields. Also, our results refer to the field-theoretic robustness of the ChSL phase against higher-order QCD corrections in the 't Hooft large-$N_c$ expansion.    

\section{Spinor solutions}

We now address the explicit spinor solution and its normalization. The rotated spinor is given by
\begin{equation}
    \psi_{s,\lambda,k} (t, \vec{x}) = e^{ikx} \Xi_{s,\lambda} (t,y,z) \ ,
\end{equation}
where $\Xi_{s,\lambda}$ is a four-component spinor encoding the time dependence, the plane wave along $y$, and the appropriate Landau-level functions. We define the following functions:
\begin{equation}
\begin{split}
    \Phi_{1,+,\lambda} = \varphi^{(+)}_{\lambda-1} (z-z_{0,s}) \ , \quad \Phi_{2,+,\lambda} = \varphi^{(+)}_{\lambda}(z-z_{0,s}) \ , \\
    \Phi_{1,-,\lambda} = \varphi^{(-)}_{\lambda}(z-z_{0,s}) \ , \quad \Phi_{2,-,\lambda} = \varphi^{(-)}_{\lambda-1}(z-z_{0,s}) \ .
\end{split}
\end{equation}
With this, we can write the $\Xi$ spinor as
\begin{equation}\label{spinor-notx}
    \Xi_{s,\lambda}(t,y,z) = \mathcal{N} e^{-i E t} e^{i p_y y} 
    \begin{pmatrix}
        F_{s1} \, \Phi_{1,s,\lambda} \\[6pt]
        F_{s2} \, \Phi_{2,s,\lambda} \\[6pt]
        H_{s1} \, \Phi_{1,s,\lambda} \\[6pt]
        H_{s2} \, \Phi_{2,s,\lambda}
    \end{pmatrix} \ . \quad
\end{equation}
In the high-density regime $\alpha'(x)=a$, substituting this solution into the Hamiltonians in Eqs.~\eqref{eq:Hplus-block} and \eqref{eq:Hminus-block} yields the lower spinor components $H_{sj}$ in terms of $F_{sj}$:
\begin{equation} \label{eq:Hs1s2}
H_{s1} = \frac{kF_{s1} - s \Omega_{s,\lambda} F_{s2}}{E + g - {\frac {sa}2}} \ , \qquad H_{s2} = \frac{-kF_{s2} - s\Omega_{s,\lambda} F_{s1}}{E + g + {\frac {sa}2}} \ .
\end{equation}
Substituting Eq.~\eqref{eq:Hs1s2} into the upper two components yields a homogeneous system of equations:
\begin{equation} \label{systemM}
\mathcal{M}_{s,\lambda }\begin{pmatrix} F_{s1} \\ F_{s2} \end{pmatrix} = 0 \ ,
\end{equation}
where the coefficient matrix is given by
\begin{equation}
\mathcal{M}_{s,\lambda(E,k)} = \begin{pmatrix} 
\epsilon_{--}\epsilon_{+-}\epsilon_{++} - k^2\epsilon_{++} - \Omega_{s,\lambda}^2\epsilon_{+-} & {ak\Omega_{s,\lambda}} \\[2mm] 
{ak\Omega_{s,\lambda}} & \epsilon_{-+}\epsilon_{+-}\epsilon_{++} - k^2\epsilon_{+-} - \Omega_{s,\lambda}^2\epsilon_{++} 
\end{pmatrix} \ ,
\end{equation}
and where we have defined\footnote{Here we have omitted the labels for brevity.}
\begin{equation}
\epsilon_{\pm \pm} = E^{(\nu)}_{s,\lambda,\eta,\zeta} \pm g \pm \frac s2 a \ .
\end{equation}
Now, we can choose the first component as $F_{s1}=1$, and set $F_{s2} =\beta^{(\nu)}_{s,\lambda,\eta,\zeta}  $. Then, we can solve the system of equations, obtaining
\begin{equation}
    \beta^{(\nu)}_{s,\lambda,\eta,\zeta} = - \frac{\epsilon_{--} \epsilon_{+-} \epsilon_{++} - k_\nu^2 \epsilon_{++} - \Omega^2_{s,\lambda} \epsilon_{+-}}{{a k_\nu \Omega_{s,\lambda}}} \ ,
\end{equation}
and then, the remaining components can be solved as follows
\begin{align}\label{eq:h-branch-app}
h^{(\nu)}_{1;s,\lambda,\eta,\zeta}\equiv  H_{s1}
&=\frac{k_\nu-s\Omega_{s,\lambda}\beta_{s,\lambda,\eta,\zeta}^{(\nu)}}
{E_{s,\lambda,\eta,\zeta}^{(\nu)}+g-\frac{s a}{2}} \ ,\\[1mm]
 h^{(\nu)}_{2;s,\lambda,\eta,\zeta} \equiv H_{s2}
&=\frac{-k_\nu\beta_{s,\lambda,\eta,\zeta}^{(\nu)}-s\Omega_{s,\lambda}}
{E_{s,\lambda,\eta,\zeta}^{(\nu)}+g+\frac{s a}{2}} \ .
\end{align}
Considering that the chiral rotation $\hat{S}$ is a unitary transformation, the probability density satisfies the relation $\psi^\dagger \psi = |\mathcal{N}^{(\nu)}|^2 \Xi^\dagger \Xi$. In the following, we assume that the boundary conditions along the $z$-coordinate are negligible for the Landau-level functions. This approximation is justified if these functions are well-localized or if the characteristic magnetic length is much smaller than the physical length of the interval. Under these assumptions, we approximate the normalization as:
\begin{equation}
\int dz \, |\Phi_{j,s,\lambda}(z-z_{0,s})|^2 \approx 1 \ ,\quad j =1,2.\end{equation}
For a box of length $L$ in the $x$-direction and length $L_y$ in the $y$-direction, the normalization condition $\int \psi^\dagger \psi \, dV = 1$, implies:
\begin{equation}
1 = |\mathcal{N}^{(\nu)}|^2 L L_y \left( |F_1|^2 + |F_2|^2 + |H_1|^2 + |H_2|^2 \right) \ ,
\end{equation}
where we have dropped the $s=\pm1$ label for simplicity.
Then,
\begin{equation}
\mathcal{N}^{(\nu)}_{s,\lambda} = \frac{1}{\sqrt{L L_y \left( |F_1|^2 + |F_2|^2 + |H_1|^2 + |H_2|^2 \right)}} \ .
\end{equation}
Using Eq. \eqref{eq:Hs1s2}, we obtain the explicit form of the normalization constant, namely
\begin{equation}
\mathcal{N}^{(\nu)}_{s,\lambda,\eta,\zeta} = \left[ L L_y \left( 1 + |\beta^{(\nu)}_{s,\lambda,\eta,\zeta}|^2 + \frac{|k_\nu - s\Omega_{s,\lambda} \ \beta^{(\nu)}_{s,\lambda,\eta,\zeta}|^2}{|E^{(\nu)}_{s,\lambda,\eta,\zeta} + g - {sa/2}|^2} + \frac{|-k_\nu \beta^{(\nu)}_{{s,\lambda,\eta,\zeta}} -s \Omega_{s,\lambda}|^2}{|E^{(\nu)}_{s,\lambda,\eta,\zeta} + g + {sa/2}|^2} \right) \right]^{-1/2} \ .
\end{equation}
Consequently, the fully normalized spinor, for $\lambda\geq1$, is given by
\begin{equation}
\Psi^{(\nu)}_{s,\lambda,\eta,\zeta}(t,x,y,z) = \mathcal{N}^{(\nu)}_{s,\lambda,\eta,\zeta} e^{-itE_{s,\lambda,\eta,\zeta}^{(\nu)}} e^{ip_y y} e^{ik_\nu x} e^{-\frac{i}{2}s\gamma_5\alpha(x)}
\begin{pmatrix}
\Phi_{1,s,\lambda} \\[2mm]
\beta^{(\nu)}_{s,\lambda,\eta,\zeta} \Phi_{2,s,\lambda}\\[2mm]
h^{(\nu)}_{1;s,\lambda,\eta,\zeta} \Phi_{1,s,\lambda}\\[2mm]
h^{(\nu)}_{2;s,\lambda,\eta,\zeta} \Phi_{2,s,\lambda}
\end{pmatrix} \ .
\end{equation}
Analogously, we obtain the normalized spinors at $\lambda=0$. Solving Eqs.~\eqref{eq:hamLLLplus} and \eqref{eq:hamLLLminus}, and setting the upper component to $1$ for convenience, yields
\begin{equation}
\Psi^{(\nu)}_{+,0,\eta,}(t,x,y,z) = \mathcal{N}^{(\nu)}_{\eta} e^{-itE_{LLL,\eta}^{(\nu)}} e^{ip_y y} e^{ik_\nu x} e^{-\frac{i}{2}\gamma_5\alpha(x)} \varphi^{(+)}_0 (z-z_{0,+})
\begin{pmatrix}
0 \\ 1\\0\\ \frac{-k_\nu}{E_{LLL,\eta}^{(\nu)}+g+ \frac a2 } \\ 0\end{pmatrix} \ ,
\end{equation}
\begin{equation}
\Psi^{(\nu)}_{-,0,\eta,}(t,x,y,z) = \mathcal{N}^{(\nu)}_{\eta} e^{-itE_{LLL,\eta}^{(\nu)}} e^{ip_y y} e^{ik_\nu x} e^{\frac{i}{2}\gamma_5\alpha(x)} \varphi^{(-)}_0 (z-z_{0,-})
\begin{pmatrix}
1 \\ 0\\ \frac{k_\nu}{E_{LLL,\eta}^{(\nu)}+g+ \frac a2 } \\ 0\end{pmatrix} \ ,
\end{equation}
with the normalization constant given by
\begin{equation}
    \mathcal{N}^{(\nu)}_{\eta}    = \left[ L L_y \left(1 + \frac{k_\nu^2}{(E_{LLL,\eta}^{(\nu)} + g + a/2)} \right)\right]^{-1/2} .
\end{equation}}

\end{document}